# Orthogonal magnetic structures of Fe$_4$O$_5$: representation analysis and DFT calculations

V.S. Zhandun[1], N.V. Kazak[1], I. Kupenko[2], D.M. Vasiukov[3], X. Li[2], E. Blackburn[3], and S.G. Ovchinnikov[1]

[1]*Kirensky Institute of Physics, Federal Research Center KSC SB RAS, 660036 Krasnoyarsk, Russia*

[2]*Institut für Mineralogie, University of Münster, Corrensstr. 24, 48149 Münster, Germany*

[3]*Division of Synchrotron Radiation Research, Department of Physics, Lund University, Box 118, Lund 221 00, Sweden*

**Abstract.** The magnetic and electronic structures of Fe$_4$O$_5$ have been investigated at ambient and high pressures via a combination of representation analysis, density functional theory (DFT+U) calculations, and Mössbauer spectroscopy. A few spin configurations corresponding to the different irreducible representations have been considered. The total-energy calculations reveal that the magnetic ground state of Fe$_4$O$_5$ corresponds to an orthogonal spin order. Depending on the magnetic propagation vector *k* two spin ordered phases with minimal energy differences are realized. The lowest energy magnetic phase is related to *k* = (0, 0, 0) and is characterized by the ferromagnetic ordering of the iron magnetic moments at prismatic sites along the *b* axis and antiferromagnetic ordering of iron moments at octahedral sites along the *c* axis. For the *k* = (1/2, 0, 0) phase, the moments in the prisms are antiferromagnetically ordered along the *b* axis and the moments in the octahedra are still antiferromagnetically ordered along the *c* axis. Under high pressure, the Fe$_4$O$_5$ exhibits magnetic transitions with corresponding electronic transitions of the metal-insulator type. At a critical pressure $P_C$ ~ 60 GPa the Fe ions at the octahedral sites undergo a high-spin to low-spin state crossover with a decrease in the unit-cell volume of ~ 4%, while the Fe ions at the prismatic sites remain in the high-spin state up to 130 GPa. This site-dependent magnetic collapse is experimentally observed in the transformation of Mössbauer spectra measured at room temperature and high pressures.

1. **INTRODUCTION**

Since the discovery of novel high-pressure iron oxides their crystal chemistry and physical properties are of great interest [1-15]. Based on the common packing motif these compounds constitute a homologous series *n*FeO·*m*Fe$_2$O$_3$, containing η-Fe$_2$O$_3$, HP-Fe$_3$O$_4$, Fe$_4$O$_5$, $Fe_5O_6$, $Fe_5O_7$, *etc*. [4]. Their crystal structures consist of two main building blocks, namely slabs of edge-shared Fe-O octahedra separated by arrays of one-dimensional (1D) chains of Fe ions coordinated by a trigonal prism (see Fig. 1). Some members of the series are recoverable at ambient conditions allowing a careful investigation of their physical properties [4, 7]. Interestingly, the majority of these oxides feature iron ions in the mixed-valent state resulting in a sizable conductivity [4, 7, 11] and intricate charge-ordering transitions at low temperatures [4,7,12], similar to the famous Verwey transition in magnetite $Fe_3O_4$ [16]. At high pressures, a site-selective spin transition is expected in these oxides [8, 17] that should have a serious effect on electronic structure [15].

Recently, a unified approach to this series based on a specific crystallographic generation mechanism was proposed which naturally classifies these oxides in terms of a sequence of the slabs or, in other words, a slab cycle [13]. Depending on the type of the slab cycle this series can be divided into two groups, an orthorhombic *N*-family and a monoclinic *N*, *N*+1 family. The former is built by only one slab of *N* octahedra wide, exemplified by $\eta$-$Fe_2O_3$, HP-$Fe_3O_4$, $Fe_4O_5$, and $Fe_5O_6$ with *N* = 1, 2, 3, and 4, respectively. The latter family has alternating slabs of two different widths, *N* and *N*+*1,* and consists of $Fe_5O_7$, $Fe_7O_9$, and $Fe_9O_{11}$ with (1, 2), (2, 3) and (3, 4) slab cycles, respectively. For the *N*-family in the charge-averaged state, a generic magnetic structure was proposed where the collinear slab and chain subsystems are orthogonal to each other [13]. The resulting magnetic structure will be either ferromagnetic or antiferromagnetic depending on whether *N* is odd or even, respectively. The remarkable feature here is that the 1D chains are *symmetry protected* from the magnetic perturbation of the surrounding slabs and order magnetically at much lower temperatures relative to the slabs, about 100 K versus 300 K [13]. These features, together with the fact that *ab initio* calculations show a strong dependence of the anisotropic conductivity on the magnetic state [13], show the great potential of these compounds to host various magnetotransport phenomena and even to realize theoretical models of coupled 1D wires [18, 19].

It is therefore important to carefully investigate this conjecture about magnetic structures in the *N*-family. In the present work we focus on $Fe_4O_5$, the *N* = 3 member of the *N*-family, to investigate its magnetic ground state. We have applied a combination of irreducible representation analysis and density-functional theory plus Hubbard U (DFT + U) calculations to analyze the possible magnetic structures at ambient pressure. For the parent space group *Cmcm* (#63), the total energies of the different spin configurations involving iron ions at all symmetry nonequivalent sites have been calculated. We find that in $Fe_4O_5$ two spin structures with *k* = (0, 0, 0) and (1/2, 0, 0) propagation vectors can be realized with a small energy difference ($\Delta E$ = 25.6 meV/unit cell). Both configurations correspond to the orthogonal structures where spins in the 1D chains are directed along the *b* axis whereas the spins in the slabs are aligned along the *c* axis. We then studied the evolution of electronic structures under high pressure. A site-dependent spin-state crossover was found, accompanied by the electronic metal-insulator transition. This result is supported by the room-temperature Mössbauer spectroscopy data which reveal a transformation of the spectral components associated with the high-spin to low-spin transition of the iron ions above 50 GPa.

## 2. METHODS

The *ab initio* calculations were performed using the Vienna *ab initio* simulation package (VASP) [20] with projector-augmented wave (PAW) pseudopotentials [21, 22]. The valence electron configuration $3d^64s^2$ was taken for the Fe atom, and $2s^22p^2$ for the O atom. The calculations are based on the density-functional theory with the Perdew-Burke-Ernzerhoff (PBE) parameterization [23] of the exchange-correlation functional and the generalized gradient approximation (GGA). Spin-orbit coupling (SOC) was turned on during all DFT-GGA calculations. The plane-wave cutoff energy was 500 eV. We used the $11 \times 4 \times 3$ Monkhorst-Pack grids of special points [24] for the Brillouin-zone integration. The energy convergence criteria were $10^{-5}$ eV and $10^{-4}$ eV for electronic and ionic relaxations, correspondingly. Structural optimizations were performed for all structures, including lattice parameters and internal coordinates. We used the GGA+U approach within the Dudarev approximation [25] and the Coulomb parameter U = 3.2 eV for the best match between experimental [3] and theoretical values of the bandgap width. The BASIREPS software [26] was used for the propagation vectors $k = (0, 0, 0)$ and $k = (1/2, 0, 0)$.

$^{57}Fe_4O_5$ were synthesized in a 1000-ton Walker-type multi-anvil apparatus at the Institut für Mineralogie at the Westfälische Wilhelms-Universität Münster (WWU), Germany from stoichiometric mixtures of fine powders of $^{57}Fe_2O_3$ and $^{57}Fe$. The syntheses were performed at pressure of 11 GPa and temperature of 1350 °C over 6 hours. We employed a standard assembly including an Au cylindrical sample capsule, a $LaCrO_3$ heater, a W3Re/W25Re thermocouple, and other components packed inside an octahedron made of Cr-bearing MgO. The synthesized crystals were preselected by single crystal diffraction at the ID15b beamline of the ESRF and Synchrotron Mössbauer Source (SMS) spectroscopy (see details below).

The high-pressure measurements were performed employing membrane-type diamond anvil cells [27]. Diamonds with 250 μm culets were employed. Compression chambers were prepared from 200 μm thick Re gaskets pre-indented to about 30 μm thickness and drilled with a hole of about 125 μm in the center of the indentation. $Fe_4O_5$

single crystals were with dimensions of about 15 × 15 × 10 μm³ were loaded into the pressure chamber. Helium gas loaded at 1.2 kbar was used as the pressure transmitting medium.

SMS spectra were acquired at the Nuclear Resonance Beamline [28] ID18 of the ESRF using the (111) nuclear reflection of a $^{57}$FeBO$_3$ single crystal mounted on a Wissel velocity transducer driven with a sinusoidal wave form [29]. The source provides $^{57}$Fe resonant radiation at 14.4 keV within a bandwidth of 15 neV (0.31 mm/s) which is tunable in energy. The X-ray beam was focused down to 13 × 4 μm² (VxH) by Kirkpatrick-Baez mirrors. The linewidth of the SMS and the absolute position of the center shift (CS) relative to α-iron were controlled before and after each measurement with a K$_2$Mg$^{57}$Fe(CN)$_6$ reference single line absorber. The velocity scale was calibrated using 25 μm thick natural iron foil. Typical collection times for each collected spectrum were about 1 hour. SMS spectra were fitted using a transmission integral with a normalized Lorentzian-squared source lineshape using the MossA software package [30].

## 3. RESULTS AND DISCUSSION

Fe$_4$O$_5$ crystallizes in an orthorhombic space group *Cmcm* with three symmetrically distinct iron sites, namely two octahedral sites at Wyckoff positions 4a and 8f (Fe1 and Fe2, respectively) and the trigonal prismatic site belonging to the position 4c (Fe3). The calculated optimized unit cell parameters are $a$ = 2.94 Å, $b$ = 10.03 Å, $c$ = 12.57 Å, and $V$ = 371 Å³, which are within 3% agreement with the experimental values reported in Ref. [1]. Three edge-sharing octahedra constitute the $N$ = 3 slab, while trigonal prisms form the 1D chains along the *a* axis (Fig. 1).

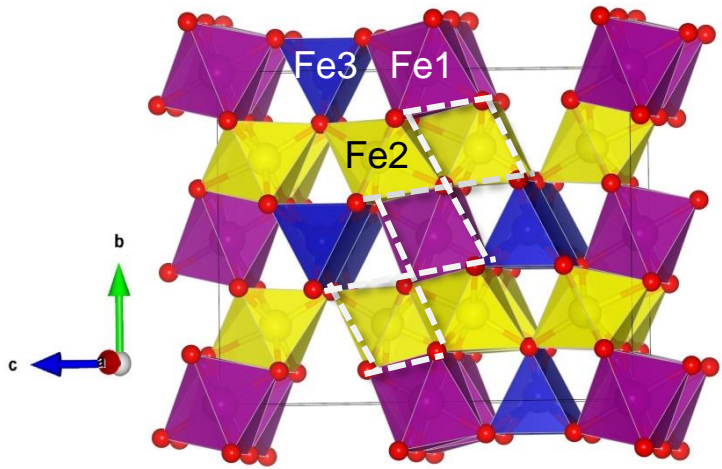

Figure 1. Crystal structure of Fe$_4$O$_5$. The symmetry distinct iron sites are highlighted in purple (Fe1), yellow (Fe2), and blue (Fe3). The $N$ = 3 wide octahedral slab is depicted by the white

dashed line. All figures of crystal and magnetic structures in this paper were generated using VESTA software [31].

The conjecture of a generic magnetic structure for the *N*-family proposed in [13] relies on the reported magnetic structures of several compounds isostructural to $Fe_4O_5$ but with the chains occupied by another divalent ion, namely Ca, Mn, Co, instead of $Fe^{2+}$ [17,32-35]. In $CaFe_3O_5$, with the non-magnetic Ca ion in the chains, phase separation into two different collinear magnetic structures was reported at 302 K, with $k = (0,0,0)$ and $k = (1/2,0,0)$ propagation vectors [33]. The subsequent detailed study [35] concluded that stochiometric $CaFe_3O_5$ adopts the $k = (1/2,0,0)$ magnetic structure while the $k = (0,0,0)$ structure is relevant to the non-stochiometric Fe-rich composition where Fe ions substitute Ca in the chains. All other $Me^{2+}Fe_3O_5$ oxides with magnetically active chains, including $Fe_4O_5$, show similar behavior with two magnetic transitions at around 300 and 100 K [3,17,32,34], corresponding to the ordering of the two subsystems. The slab spin system orders slightly above room temperature while the chains order independently at much lower temperatures. Notably, the neutron diffraction experiments [32, 34] revealed an *identical magnetic structures of the slabs* in all of these compounds, the same as the $k = (0,0,0)$ structure in $CaFe_3O_5$ [33]. This implies a common formation mechanism of the magnetic ground state, imposed by the crystal and magnetic symmetries in these compounds.

To determine the possible magnetic structures for $Fe_4O_5$ an irreducible representation analysis was carried out. Any axial vector configurations in periodic crystals should be classified according to the crystal symmetry and described as a certain linear combination of the basis vectors of the irreducible (axial) representations of the parent space group. Usually one or a few irreducible representations are sufficient to describe a magnetic structure in real materials. Because of aforesaid data we limited our consideration to magnetic structures with the propagation vectors $k = (0, 0, 0)$ and $k = (1/2, 0, 0)$. The decompositions on irreducible representations *(IrReps)* are presented in Table 1. According to the Landau theory, we consider only magnetic structures corresponding to the single irreducible representations, namely: $m\Gamma_2^+$, $m\Gamma_4^+$ and $m\Gamma_3^+$ for $k = (0, 0, 0)$ and $m\Sigma_1$, $m\Sigma_2$, $m\Sigma_3$, $m\Sigma_4$ for $k = (1/2, 0, 0)$. The irreducible representations and basis vectors for the spins at the 4a, 8f, and 4c sites are shown in Table 2.

Both collinear and orthogonal spin configurations can be realized. For the propagation vector $k = (0, 0, 0)$ the basis vectors of *IrRep* $m\Gamma_3^+$ correspond to the collinear spin configurations only, whereas the basis vectors of $m\Gamma_2^+$ and $m\Gamma_4^+$ representations correspond to both collinear and non-collinear orthogonal configurations with the Fe magnetic moments in slabs lying in *bc* plane and Fe magnetic moments in prisms ordered along *b*- or *c*-axis. The collinear spin configurations correspond to the ferromagnetic or ferrimagnetic ordering of all Fe magnetic moments directed along the *a*- (*IrRep* m$\Gamma_3^+$), *b*- (*IrRep* m$\Gamma_4^+$, u=0, v≠ 0) and *c*-axis (*IrRep* m$\Gamma_2^+$, u=0, v≠ 0). The orthogonal spin configurations result from the decomposition of the initial Fe magnetic system into two subsystems. The first subsystem, the slabs, consists of the magnetic ions in the octahedral coordination (4a and 8f), while the second one, the chains, is made up of the Fe ions in the trigonal prisms (4c). In the slabs, there is an antiferromagnetic alignment of spins along the *b*-axis (m$\Gamma_2^+$) or *c*-axis (m$\Gamma_4^+$) with the ferromagnetic component along the *c*- or *b*-axis, respectively, suggesting a canted spin structure. The magnetic moments in the prisms are ferromagnetically ordered along *b*- or *c*-axis and orthogonal to the corresponding antiferromagnetic component of magnetic moments in the slabs. For the *k* = (1/2, 0, 0) phase various collinear and non-collinear spin configurations are possible. Here, following the neutron diffraction data for isostructural compounds MnFe$_3$O$_5$ and CoFe$_3$O$_5$ [32, 34], we consider only spin configurations with magnetic moments in the 4c positions ordered orthogonal to the *c* axis.

Table 1. The decompositions on irreducible representations with magnetic propagation vectors *k* = (0, 0, 0) and *k* = (1/2, 0, 0) for the space group *Cmcm* (#63).

|  | *k* = (0, 0, 0) | *k* = (1/2, 0, 0) |
|---|---|---|
| Γ(4a) | m$\Gamma_1^+$ ⊕ 2·m$\Gamma_2^+$ ⊕ 2·m$\Gamma_4^+$ ⊕ m$\Gamma_3^+$ | m$\Sigma_1$ ⊕ m$\Sigma_2$ ⊕ 2·m$\Sigma_4$ ⊕ 2·m$\Sigma_3$ |
| Γ(8f) | m$\Gamma_1^+$ ⊕ 2·m$\Gamma_1^-$ ⊕ 2·m$\Gamma_2^+$ ⊕ m$\Gamma_2^-$ ⊕ m$\Gamma_3^+$ ⊕ 2·m$\Gamma_3^-$ ⊕ 2·m$\Gamma_4^+$ ⊕ m$\Gamma_4^-$ | 3·m$\Sigma_1$ ⊕ 3·m$\Sigma_2$ ⊕ 3·m$\Sigma_4$ ⊕ 3·m$\Sigma_3$ |
| Γ(4c) | m$\Gamma_1^-$ ⊕ m$\Gamma_2^+$ ⊕ m$\Gamma_2^-$ ⊕ m$\Gamma_3^+$ ⊕ m$\Gamma_3^-$ ⊕ m$\Gamma_4^+$ | m$\Sigma_1$ ⊕ 2·m$\Sigma_2$ ⊕ m$\Sigma_4$ ⊕ 2·m$\Sigma_3$ |

Table 2. Irreducible representations (*IrReps*)/magnetic space groups (*MSG*) and basis vectors for Fe(4a), Fe(8f), and Fe(4c) spin order in Fe$_4$O$_5$ with propagation vectors *k* = (0, 0 ,0) and (1/2, 0, 0). The magnetically independent atoms Fe$_i$, where *i*=1-8, are given only. (For more information on the magnetic atoms numbering see Supplementary Materials, Figure S1).

| Atom numbers | Fe1 | Fe2 | Fe3 | Fe4 | Fe5 | Fe6 | Fe7 | Fe8 |
|---|---|---|---|---|---|---|---|---|
| Iron sites | Octahedron | | Octahedron | | | | Tr. prism | |
| Wyckoff positions | 4a | 4a | 8f | 8f | 8f | 8f | 4c | 4c |
| *IrReps*/*MSG* | *k* = (0, 0, 0) | | | | | | | |
| m$\Gamma_2^+$/*Cm'c'm* (#63.462) | (0uv) | (0-uv) | (0uv) | (0-uv) | (0-uv) | (0uv) | (00u) | (00u) |
| m$\Gamma_4^+$/*Cm'cm'* | (0uv) | (0u-v) | (0uv) | (0u-v) | (0u-v) | (0uv) | (0u0) | (0u0) |
| *(#63.464)* | | | | | | | | |
| *mΓ₃⁺/Cmc'm'* *(#63.463)* | (u00) | (u00) | (u00) | (u00) | (u00) | (u00) | (u00) | (u00) |
| | \multicolumn{8}{c}{$k = (1/2, 0, 0)$} |
| *mΣ₁/P$_a$bcm* *(57.386)* | (u,0,0) | (-u,0,0) | (u,v,w) | (u,-v,-w) | (-u,-v,w) | (-u,v,-w) | (0,0,u) | (0,0,-u) |
| *mΣ₂/P$_b$bcn* *(60.427)* | (u,0,0) | (u,0,0) | (u,v,w) | (u,-v,-w) | (u,v,-w) | (u,-v,w) | (u,v,0) | (u,-v,0) |
| *mΣ₄/P$_c$nma* *(62.452)* | (0,u,v) | (0,-u,v) | (u,v,w) | (-u,v,w) | (-u,-v,w) | (u,-v,w) | (0,0,u) | (0,0,u) |
| *mΣ₃/P$_a$bca* *(61.438)* | (0,u,v) | (0,u,-v) | (u,v,w) | (-u,v,w) | (u,v,-w) | (-u,v,-w) | (u,v,0) | (-u,v,0) |

Based on the above representation analysis nine magnetic structures were chosen and their total energies were calculated using the DFT+U. Six of the structures (FM1-FM3, AFM1-AFM3) are related with the $m\Gamma_2^+$, $m\Gamma_4^+$ and $m\Gamma_3^+$ irreducible representations for $k = (0, 0, 0)$ and the AFM4, AFM5 and AFM6 configurations correspond to the $m\Sigma_1$, $m\Sigma_4$ and $m\Sigma_3$ irreducible representations for $k = (1/2, 0, 0)$, respectively (Table 3). All considered magnetic structures are plotted in Figure S2 of the Supplementary Materials. The orthogonal FM3 structure (*IrRep* $m\Gamma_4^+$), presented in Figure 2a, was found to have the lowest energy. In this phase the magnetic moments in the chains are aligned along the *b* axis and have a ferromagnetic *interchain* order. Our calculations therefore support the magnetic structure proposed for $Fe_4O_5$ in Ref. [13] while the magnetic structure reported in Ref. [3] has a much higher energy by 10.4 meV per Fe atom (167.3 meV/cell).

Interestingly, the second most energetically favorable configuration is the orthogonal structure AFM6 (*IrRep* $m\Sigma_3$) with antiferromagnetic *interchain* order as shown in Figure 2b. Although the magnetic moments in the slabs are ordered along *c*-axis in both phases, there is a crucial difference. The FM3 phase has an antiferromagnetic spin alignment within the triads Fe2(↑)-Fe1(↓)-Fe2(↑) (Fig. 2a), whereas AFM6 phase has the ferromagnetic order Fe2(↑)-Fe1(↑)-Fe2(↑) (Fig. 2b). The difference in triads actually reflects the fact that AFM6 corresponds to the Type IV klassengleiche magnetic space group, which forbids any ferromagnetism, while FM3 is the Type III translationgleiche group, which allows ferromagnetic ordering.

We stress that both the FM3 and AFM6 magnetic structures of the slabs were found in $CaFe_3O_5$ from the neutron diffraction measurements at low temperatures [33]. Cassidy *et al.* [35] demonstrated that AFM6 is the true ground state in $CaFe_3O_5$. We have carried out calculations of these two magnetic structures in $CaFe_3O_5$ and indeed, the AFM6 is lower in energy by ~ 6.5 meV per Fe atom (or 78.4 meV per unit cell) than the FM3 structure (Table S1 in the Supplementary Materials). In $Fe_4O_5$ the situation is the opposite, and FM3 is lowest energy state, although the energy difference between these orthogonal phases is rather small (~ 1.6 meV per Fe atom or ~25.6 meV per unit cell, Table 3). Apparently, the difference is due to the magnetic interactions in the chains. In $CaFe_3O_5$ the chains are occupied by non-magnetic Ca, whereas for $Fe_4O_5$ ions should have a ferromagnetic *intrachain* interactions that favors the FM3 structure with $k = (0, 0, 0)$ (see Ref. [13] for more details).

Another consequence of the magnetically active chains is the small canting of magnetic moments in the slabs towards the $b$ axis obtained in the calculations (Fig. 2). This can be related to the influence of the ordered 1D chains because only the chains are symmetry-protected against perturbations from the slabs but not vice versa. In our $CaFe_3O_5$ calculations do not find any sign of such canting, see Figure S3 and Table S1 in the Supplementary Materials. One can conclude that the filling of the prismatic sites by the magnetic ion causes the disturbance of the collinear antiferromagnetic order in the slabs and appearance a small canting of the magnetic Fe moments towards $b$-axis. We also note that our calculations were carried out for the charge-averaged structure. Since charge ordering decreases the symmetry of crystal structure [3, 7] it may also affect the ground state magnetic structure. Nevertheless, it can be treated as a perturbation to the magnetic structures studied here and shouldn't affect the orthogonal nature of the spin configurations.

Table 3. The spin configurations and total energies corresponding to the irreducible representations (*IrReps*) for the Fe(4a), Fe(8f), and Fe(4c) spin order in $Fe_4O_5$ with propagation vectors (0, 0, 0) and (1/2, 0, 0). The ↑ stands for spin up and ↓ for spin down, and the order of signs corresponds to the labels of the Fe ions. The orthogonal spin configuration is denoted by 'orth'. The energy is given relative to the energy of the FM3 magnetic configuration.

| *IrReps* | Magnetic ordering | Magnetic propagation vector, $k$ | Spin config. | ΔE (meV/Fe atom) |
|---|---|---|---|---|
| $m\Gamma_3^+$ | FM1 | (0, 0, 0) | ↑↑↑ | 29.6 |
| $m\Gamma_3^+$ | AFM1 | (0, 0, 0) | ↑↓↑ | 7.0 |
| $m\Gamma_3^+$ | AFM2 | (0, 0, 0) | ↑↓↓ | 12.1 |
| $m\Gamma_3^+$ | AFM3 | (0, 0, 0) | ↑↑↓ | 17.6 |
| $m\Gamma_2^+$ | FM2 | (0, 0, 0) | orth | 3.2 |
| $m\Gamma_4^+$ | FM3 | (0, 0, 0) | orth | 0.0 |
| $m\Sigma_1$ | AFM4 | (1/2, 0, 0) | orth | 19.9 |
| $m\Sigma_4$ | AFM5 | (1/2, 0, 0) | orth | 11.5 |
| $m\Sigma_3$ | AFM6 | (1/2, 0, 0) | orth | 1.6 |

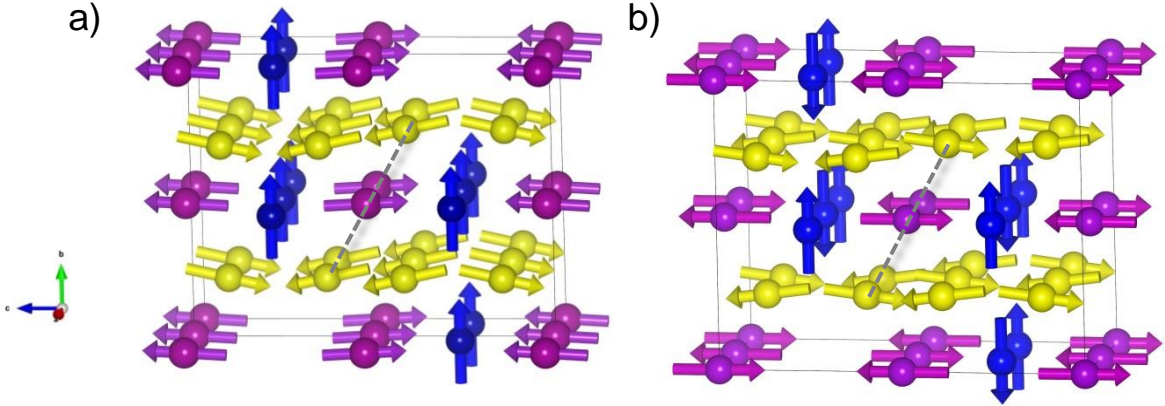

Figure 2. Orthogonal magnetic structures for the (a) FM3 and (b) AFM6 phases. The magnetic structure of the FM3 phase is shown within a unit cell doubled along the *a* axis. The arrows show the directions of the Fe spins at the Fe1 (purple), Fe2 (yellow), and Fe3 (blue) sites. The dashed lines show the (a) antiferromagnetic and (b) ferromagnetic orderings within the slab triad Fe2-Fe1-Fe2.

Table 3. The Fe occupation number ($n_d$) and the magnetic moment components for the FM3 and AFM6 spin configurations.

| Fe-site | $n_d$ | Magn. moment ($\mu_B$) | |
|---|---|---|---|
| | | FM3 | AFM6 |
| Fe1 (4a) | 5.84 | (0.0, +0.2, ±3.8) | (0.0, 0.0, ±3.8) |
| Fe2 (8f) | 5.65 | (0.0, -0.9, ±3.7) | (0.0, -0.5, ±3.9) |
| Fe3 (4c) | 5.96 | (0.0, +3.6, 0.0) | (0.0, +3.6, 0.0) |

Our calculations show that the crystal structure of $Fe_4O_5$ keeps the same orthorhombic space group *Cmcm* over the entire considered pressure range. The mean calculated bond lengths at ambient pressure are <Fe1 – O> = 2.13 Å, <Fe2 – O> = 2.07 Å, and <Fe3 – O> = 2.22 Å, pointing to a preference for the Fe3 trigonal prismatic site for $Fe^{2+}$ ions with their larger ionic radius. For the Fe1 and Fe2 sites, the calculations are consistent with the octahedral sites being occupied by iron ions in the mixed-valent state, in accordance with experimental data [3,13].

Upon increasing the pressure, the lattice parameters and unit cell volume shrink in a good agreement with experimental observations (Fig. 3). It was found that $Fe_4O_5$ undergoes a sharp volume contraction with $\Delta V/V = 4\ \%$ at around 60 GPa in accordance with pressure data reported in Refs. [1, 14, 15]. The bond length pressure dependencies also show changes in the slopes around 60 GPa (Fig. 4). The bond length difference between octahedral Fe1 and Fe2 sites is about 3% at ambient pressure and becomes ~2 % at 100 GPa. The Fe3 site has the largest bond length at all pressures. Above 60 GPa the pressure dependence of the <Fe3 – O> bond length is rather different from the octahedral sites revealing a high stiffness. The incompressibility of the trigonal prisms leads to a much slower decrease in the *a*- lattice constant (Fig. 4a).

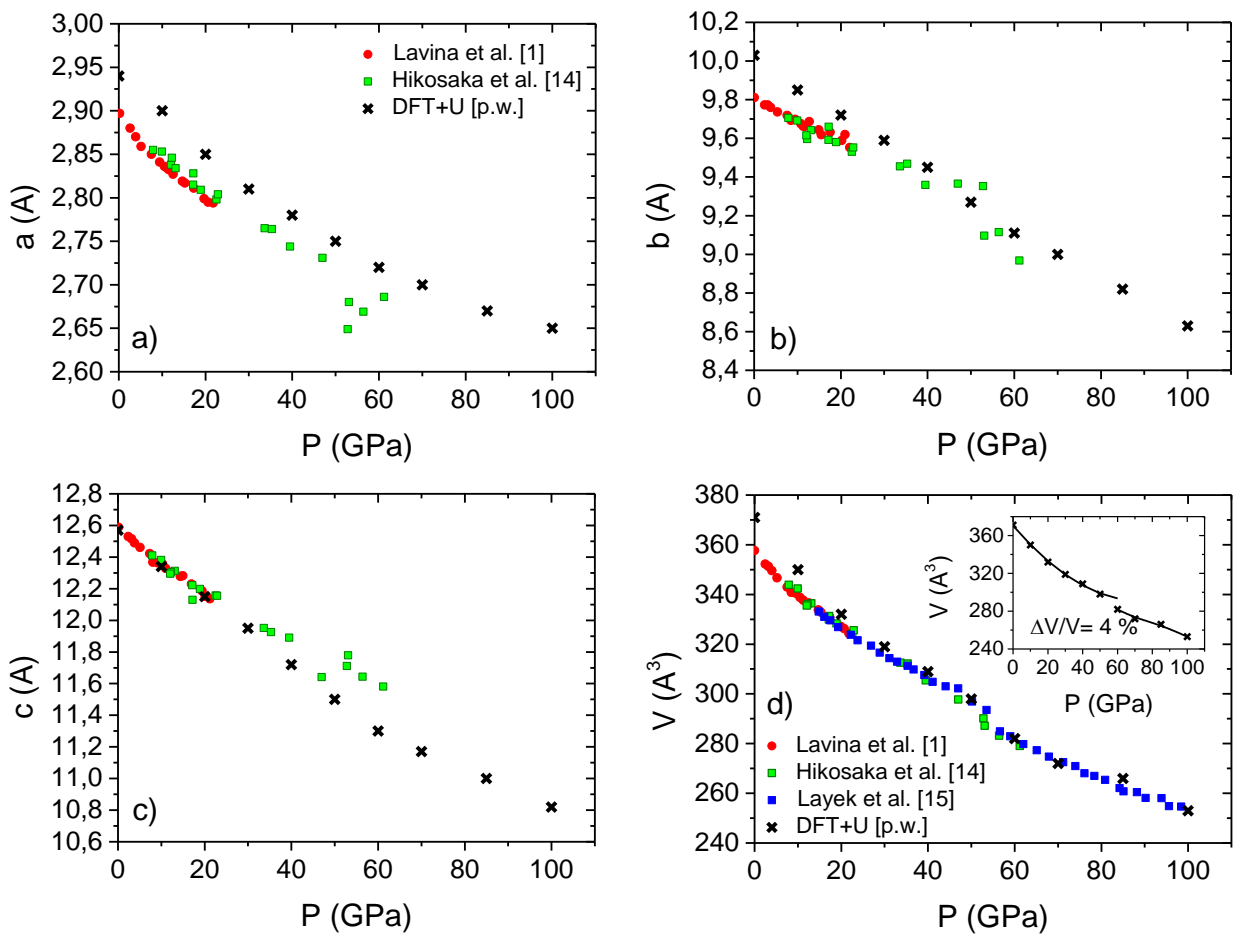

Figure 3. Pressure dependence of the lattice parameters (a)-(c) and unit cell volume (d) for $Fe_4O_5$. The red circles, green and blue squares denote the experimental data taken from Refs [1], [14], and [15] respectively. The black crosses are the calculation results. The inset shows the volume contraction near 60 GPa.

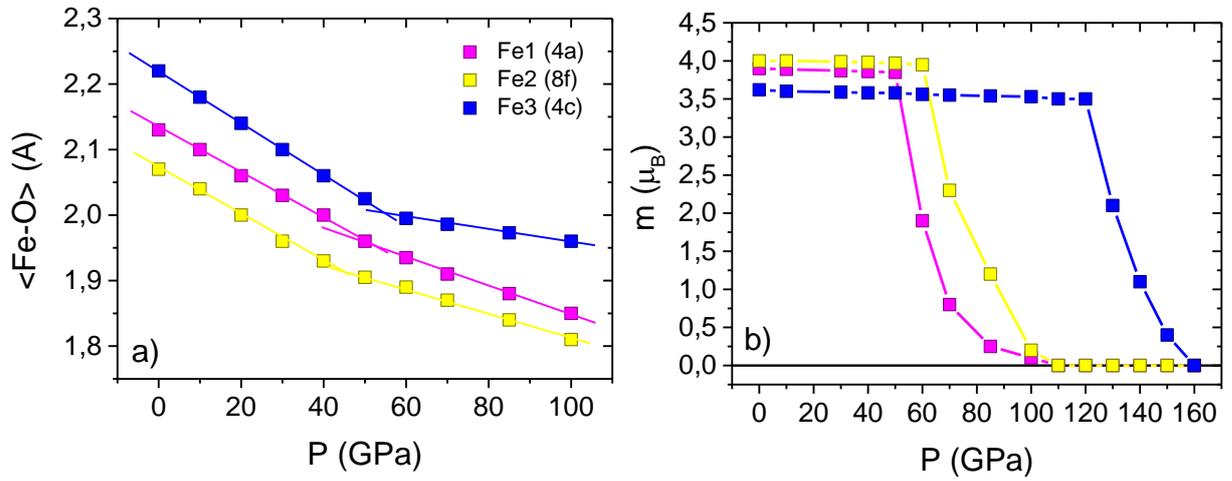

Figure 4. a) The pressure dependence of the mean bond lengths <Fe-O> at symmetry distinct iron sites. The straight lines are guide for the eye. b) Evolution of the Fe magnetic moments with pressure.

The pressure evolution of the iron magnetic moments in $Fe_4O_5$ is shown in Figure 4b. Their behavior is almost pressure-independent up to a critical pressure that is different for each iron site. $P_{C1,2,3}$ = 60, 70, and 130 GPa for the Fe1, Fe2, and Fe3 sites respectively. At $P_{C1}$ = 60 GPa the calculated magnetic moment of the Fe1 site undergoes a high-spin to low-spin crossover with a drastic reduction by ~50 %. Then, at $P_{C2}$ = 70 GPa, the magnetic moment at the octahedral Fe2 site shows a similar reduction. The collapse of the magnetic moment at the prismatic Fe3 site follows only at $P_{C3}$ = 130 GPa. This is rather important since the magnetic moments in the slabs undergo a high spin (HS)-low spin (LS) state crossover while the iron ions in the prisms remain in the HS state up to higher pressures. Three consecutive pressure-induced HS-LS crossovers, at the octahedral sites at ~58 GPa and ~80 GPa and then at the prismatic sites at ~150 GPa were obtained within the spin-polarized DFT electronic structure calculation and DFT+DMFT ones [8, 15]. The theory steadily predicts a site-dependent magnetic collapse under high pressures.

In order to clarify the electronic and magnetic states of iron ions in $Fe_4O_5$ at high pressures, we carried out a Mössbauer spectroscopy experiment at room temperature and pressures up to 54.5 GPa using the Synchrotron Mössbauer Source [29]. The hyperfine parameters of all the components are listed in Table 4. The Mössbauer spectrum at 22 GPa can be fitted by the superposition of a magnetic sextet and a paramagnetic doublet with relative areas of ~ 83 % and ~17 %, respectively (Fig.5a). For the sextet, the center shift of $\delta_{CS}$ = 0.61 mm/s corresponds to mixed-valent high-spin Fe ions in the octahedra, while for the doublet the large values of $\delta_{CS}$ = 1.019 mm/s and $\Delta E_Q$ = 2.7 mm/s is unambiguously assigned to the high-spin $Fe^{2+}$ ions in the trigonal prism. There is a large broadening of the sextet lines resulting from electron exchange between iron ions in slabs. The high width of the doublet likely indicate the onset of magnetic ordering in trigonal prisms, yet the magnitude of the hyperfine magnetic field is yet too low for its proper determination.

At $P$ = 54.5 GPa (Fig.5b) the spectrum is qualitatively different and can be fitted by a superposition of four components: two magnetic sextets, doublet and singlet with relative areas of ~20.9 %, 64.7 %, 9.6 % and 4.8 %, respectively. The sextet with $\delta_{CS}$ = 0.60 mm/s and $B_{hf}$ = 34.1 T can still be assigned to magnetically ordered mixed-valent Fe ions in octahedral coordination while another sextet with $\delta_{CS}$ =1.09 mm/s and $B_{hf}$ = 43.5 T corresponds to $Fe^{2+}$ ions in trigonal prisms. Interestingly, both doublet and singlet components have rather low center shift values of −0.30 mm/s and 0.34 mm/s, respectively, unambiguously identifying them as low-spin state iron ions [4, 36-39]. We therefore conclude from the Mössbauer data that at pressures above 50 GPa the transition of iron ions from high- to low-spin state has indeed begun, in agreement with theoretical calculations. However, the current dataset is not sufficient to reliably disentangle the behavior at the individual iron sites. These experiments are continuing and their results will be published later elsewhere.

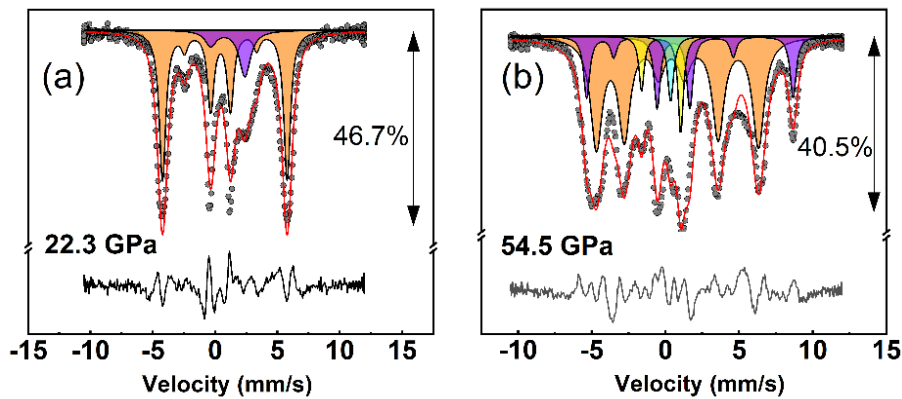

Figure 5. Room temperature Mössbauer spectra of $Fe_4O_5$ at 22.3 and 54.5 GPa.

Table 4. Hyperfine parameters of the $Fe_4O_5$ at room temperature and P=22.3, and 54.5 GPa. The $\delta_{CS}$ is the central shift (±0.01 mm/s), $\Delta E_Q/2\varepsilon$ is the quadrupole splitting/quadrupole shift (±0.02 mm/s), $B_{hf}$ is the hyperfine magnetic field (±0.1 T), FWHM is line widths (±0.09 mm/s), and A is an iron occupation factor (relative area, ±3 %).

| P (GPa) | Fe sites | $\delta_{CS}$ (mm/s) | $\Delta E_Q/2\varepsilon$ (mm/s) | $B_{hf}$ (T) | FWHM (mm/s) | A (%) |
|---|---|---|---|---|---|---|
| 22.3 | Tr. prism ($Fe^{2+}$) | 1.02 | 2.71 | | 1.08 | 17.5 |
| | Oct. ($Fe^{2+}/Fe^{3+}$) | 0.61 | 0.34 | 31.0 | 0.64 | 82.5 |
| 54.5 | Tr. prism ($Fe^{2+}$) | 1.09 | 1.13 | 43.5 | 0.49 | 20.9 |
| | Oct. ($Fe^{2+}/Fe^{3+}$) | 0.60 | 0.42 | 34.1 | 0.96 | 64.7 |
| | Oct. $Fe^{2+}$ | 0.34 | | | 0.56 | 4.8 |
| | Oct. $Fe^{3+}$ | -0.30 | 2.66 | | 0.56 | 9.6 |

Finally, the DFT+U calculations reveal that $Fe_4O_5$ undergoes a pressure-induced electronic transition. The total density of states (DOS) plots at ambient pressure and high-pressure are shown in Fig. 6. There is a strong hybridization of $d$-electrons of Fe atoms with $p$-electrons of oxygen in a wide energy interval, which increases with pressure (see projected DOS in Fig.S2 of Supplementary Materials). At ambient pressure the main contribution to the valence band near the Fermi energy is due to the $d$-electrons of the Fe1 and Fe3 ions with a small admixture of the oxygen $p$-electrons. The Fe3 $d$-electrons form a large narrow peak near the Fermi level and give a main contribution to the conductivity at the non-zero temperature. The Fe2 $d$-electrons give the largest contribution to the minority spin states of the conductive band forming the large localized peak. Thus, the ambient-pressure bandgap is mainly determined by electron transitions from occupied majority spin $d$-states of Fe1 ion to empty majority spin $d$-states of the Fe3 ion (without spin flip) and minority spin $d$-states of the Fe2 ion (with spin flip). One may see that at ambient pressure the system is insulating with a bandgap of $E_g$ =0.23 eV (Fig. 6a). At 60 GPa the Fe2 and especially Fe1 $d$-state densities undergo a significant transformation. A large broadening and non-zero density of states at the Fermi level appear. The bandgap gradually decreases and then collapses at $P_{MIT}$ = 60 GPa, which coincides with the onset of the HS-LS crossover at the octahedral sites (Figure 7). Our DFT + U calculation for the high-pressure phase shows that the insulating state does disappear and the system becomes metallic (Fig. 6b). One can conclude that the $Fe_4O_5$ undergoes a gradual metal-insulator transition. The calculated onset of the metal-insulator transition at $P_{MIT}$ is in good agreement with high-pressure resistivity measurements which show a metallization occurring at ~88 GPa [15]. Recent DMFT calculations gave a value of $P_{MIT}$ = 84 GPa [15]. The difference with our results is obviously due to the U parameter value, which is used in our spin-polarized DFT calculations. The larger the U, the more pressure is required to reach the collapse of the energy gap. We have used the value of U = 3.2 eV, which gives reasonably good agreement of the calculated bandgap with the experimental value of 0.226 eV [3].

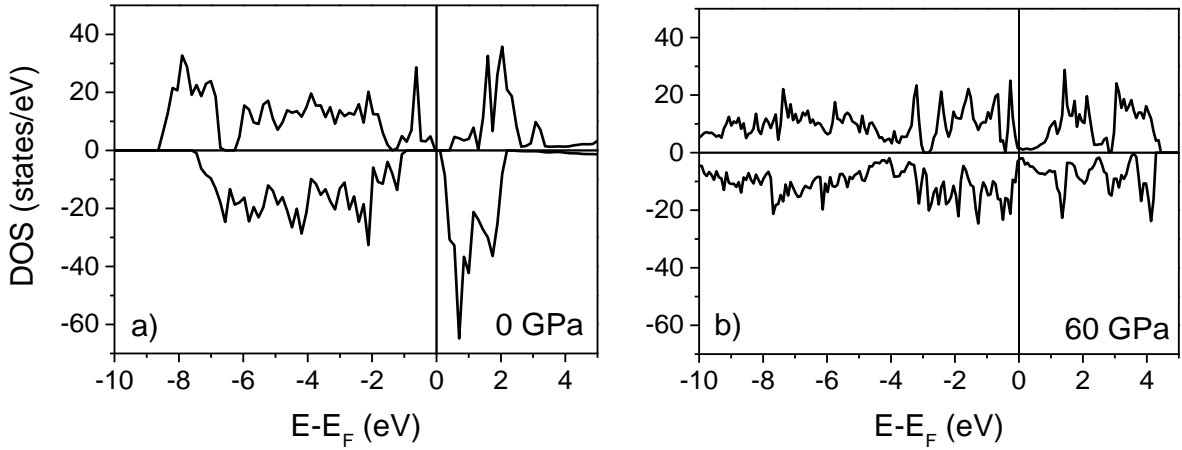

Figure 6. The total density of states (DOS) of $Fe_4O_5$ at ambient pressure (a) and $P_{MIT}$=60 GPa (b). The straight lines correspond to Fermi energy. Negative values of the DOS show spin-down states.

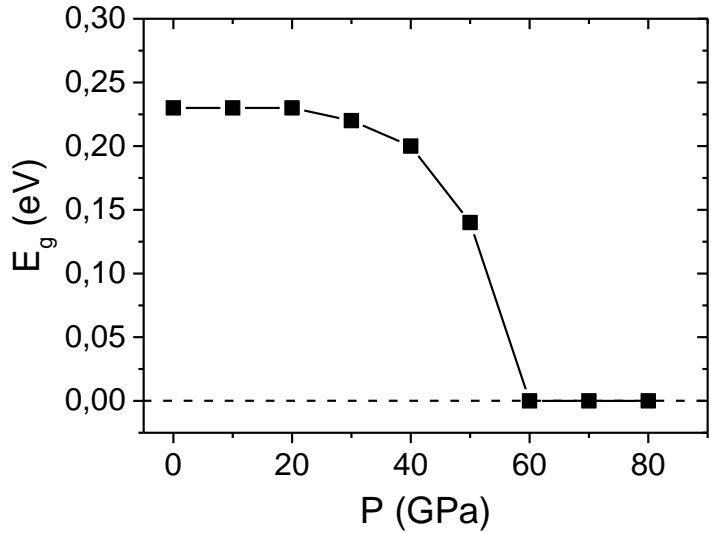

Figure 7. Evolution of the bandgap of $Fe_4O_5$ with pressure.

**CONCLUSIONS**

Our combined representation analysis and DFT+U calculations suggest that the magnetic ground state of $Fe_4O_5$ corresponds to the orthogonal magnetic structure with $k = (0, 0, 0)$ (FM3 spin configuration) with iron magnetic moments in the prisms ferromagnetically ordered along the *b* axis and moments in the slabs antiferromagnetically ordered along the *c* axis. As result, the Fe ions in the prism form the ferromagnetic 1D chains propagating along the *a* axis. The second lowest energy magnetic structure with $k = (1/2, 0, 0)$ (AFM6 spin configuration) is also orthogonal and corresponds to an antiferromagnetic ordering of the magnetic moments at the prisms along the *b* axis. Thus, for the $Fe_4O_5$ case, our results confirm the conjecture proposed in the work [13]. In both magnetic structures the chain subsystem is symmetry-protected from the slabs leading to the independent ordering of the slab and chain subsystems. However, the opposite is not true that results in the small canting of the moments in the slabs towards the *b* axis due to the influence of ordered chains.

Our calculations predict that Fe ions in the slabs undergo a HS-LS state crossover at $P_C \approx$ 60 GPa, whereas the Fe ions in the prisms remain in the HS state up to P~130 GPa. This result is supported by the room-temperature Mössbauer spectroscopy data at high pressures, where the appearance of the paramagnetic doublet and singlet was found to be related with the spin-state crossover of the mixed-valent $Fe^{3+}/Fe^{2+}$ ions in slabs. The spin crossover in the slabs should be accompanied by the metal-insulator transition according to the DFT+U calculations.

**ACKNOWLEDGMENTS**

V.S. Zhandun, N.V. Kazak and S.G. Ovchinnikov acknowledge support provided by the Russian Foundation for Basic Research (project no. 21-52-12033). I. Kupenko acknowledge the support provided by the German Research Foundation (DFG) through the DFG Project AOBJ: 674300 GZ.:KU 3832/3-1.

The calculations were performed with the computer resources of "Complex modeling and data processing research installations of mega-class" SRC "Kurchatovsky Institute" (http://ckp.urcki.ru).

We thank Arno Rohrbach and Stephan Klemme (Universität Münster) for their help with the synthesis of the $Fe_4O_5$ crystals. The authors acknowledge the European Synchrotron Radiation Facility for provision of synchrotron radiation facilities and we would like to thank G. Aprilis for assistance conduction SMS experiments and in using Nuclear Resonance beamline ID18 and D. Comboni for assistance conduction XRD experiments and using High Pressure Diffraction Beamline ID15b.

Supplementary Materials for:

Orthogonal magnetic structures of $Fe_4O_5$:

representation analysis and DFT calculations

V.S. Zhandun[1], N.V. Kazak[1], I. Kupenko[2], D.M. Vasiukov[3], X. Li[2], E. Blackburn[3], and S.G. Ovchinnikov[1]

[1]*Kirensky Institute of Physics, Federal Research Center KSC SB RAS, 660036 Krasnoyarsk, Russia*

[2]*Institut für Mineralogie, University of Münster, Corrensstr. 24, 48149 Münster, Germany*

[3]*Division of Synchrotron Radiation Research, Department of Physics, Lund University, Box 118, Lund 221 00, Sweden*

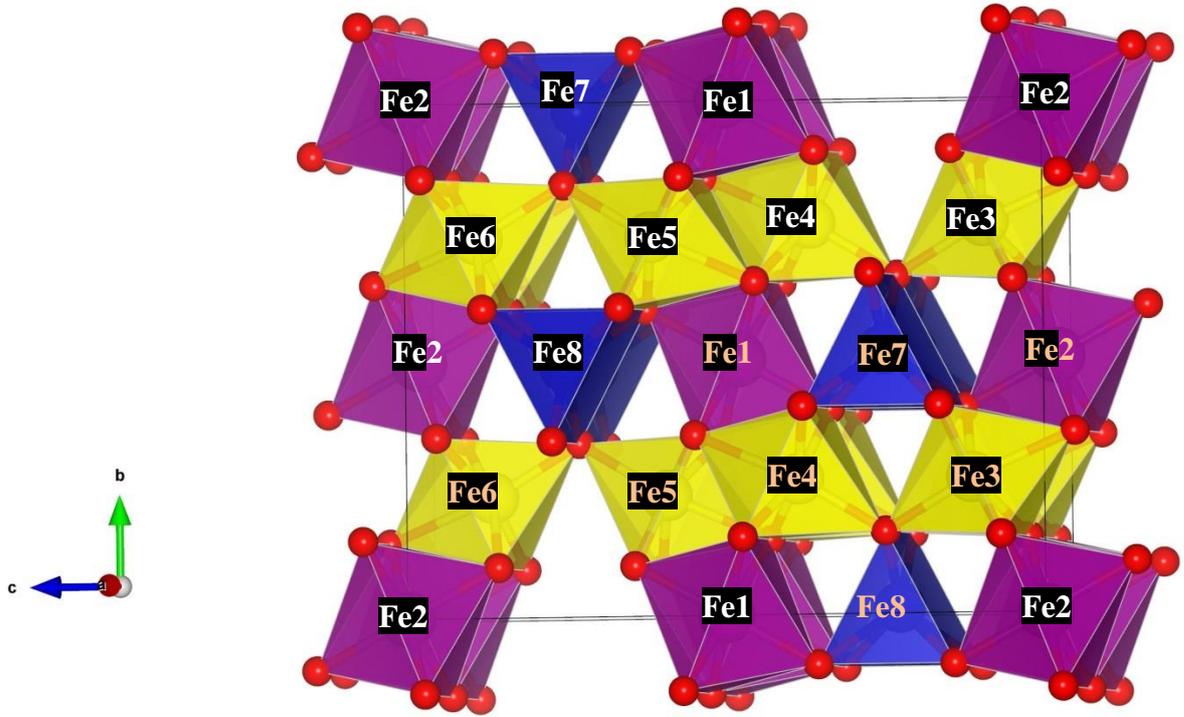

Figure S1. The numbering of the magnetically independent atoms Fe$i$ ($i$=1-8) for the magnetic propagation vector $k$ = (0, 0, 0). For $k$ = (1/2, 0, 0) the magnetic moment of $j$'-th atom is transformed according to the expression $m(R_j + a) = -\Psi(R_j)$, where $\Psi(R_j)$ is a basis vector of $i$-th Fe atom and $a$ is a lattice constant, $R_{j'} = R_j + a$.

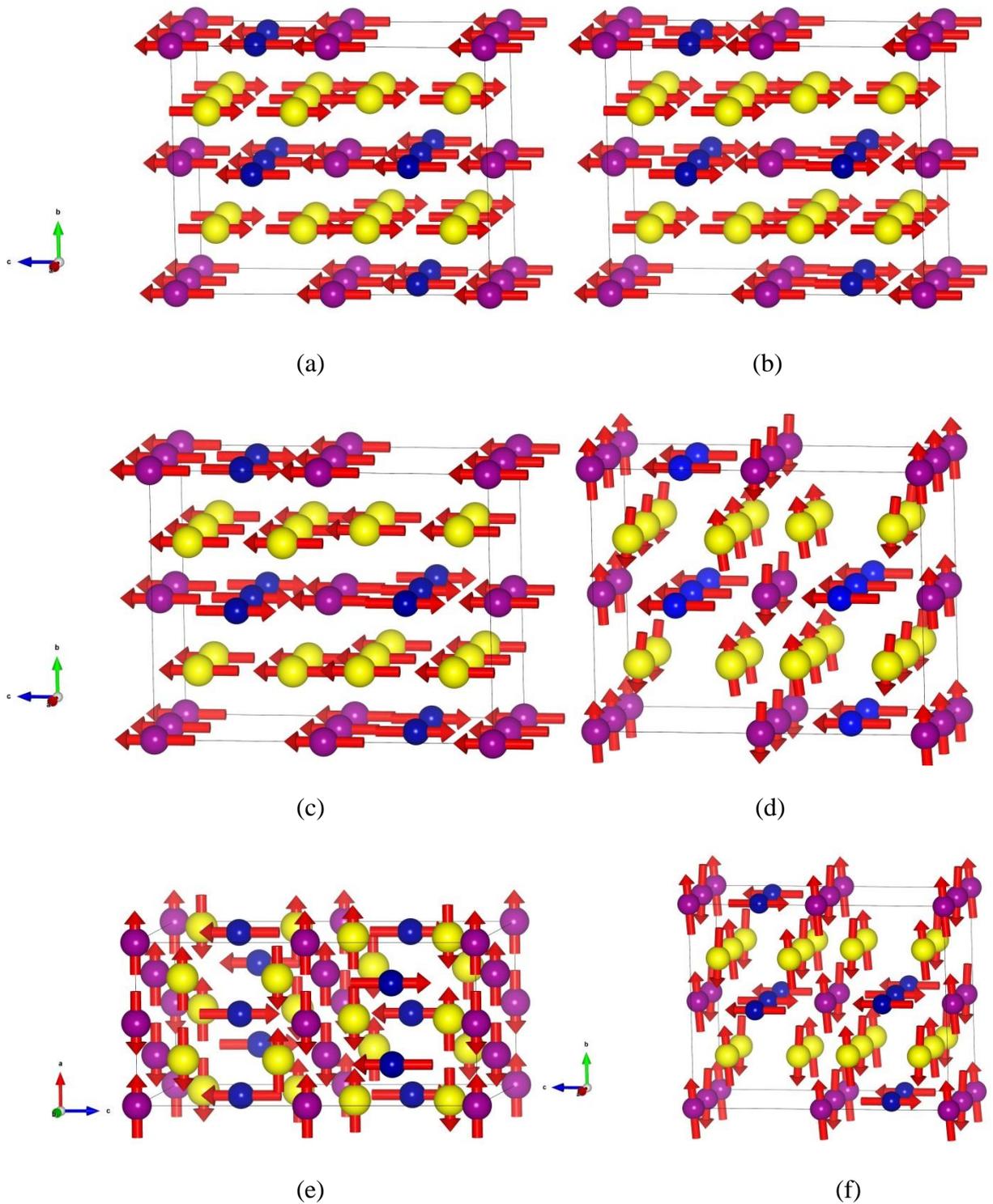

Figure S2. The AFM1-AFM3 (a-c), FM2 (d) and AFM4-5 (e-f) magnetic structures. All magnetic structures are shown within doubled unit cell along the *a* axis. The arrows show the directions of the Fe spins at the Fe1 (purple), Fe2 (yellow), and Fe3 (blue) sites.

Table S1. The spin configurations, Fe magnetic moment components, and total energies for the AFM6 and FM3 phases in CaFe$_3$O$_5$. The arrows show the spin alignment within Fe2-Fe1-Fe2 triad, the ↑ stands for spin up and ↓ for spin down.

| Magnetic (charge) ordering | Magnetic propagation vector, $k$ | Spin config. | ΔE (meV/Fe atom) | Magn. moment ($\mu_B$) | |
|---|---|---|---|---|---|
| | | | | Fe1(4a) | Fe2(8f) |
| AFM6 | (1/2, 0, 0) | ↑↓↑ | 0 | (0.00, 0.00, ±3.67) | (0.00, 0.00, ±4.10) |
| FM3 | (0, 0, 0) | ↑↑↑ | 6.5 | (0.00, 0.00, ±3.60) | 0.00, 0.00, ±4.15) |

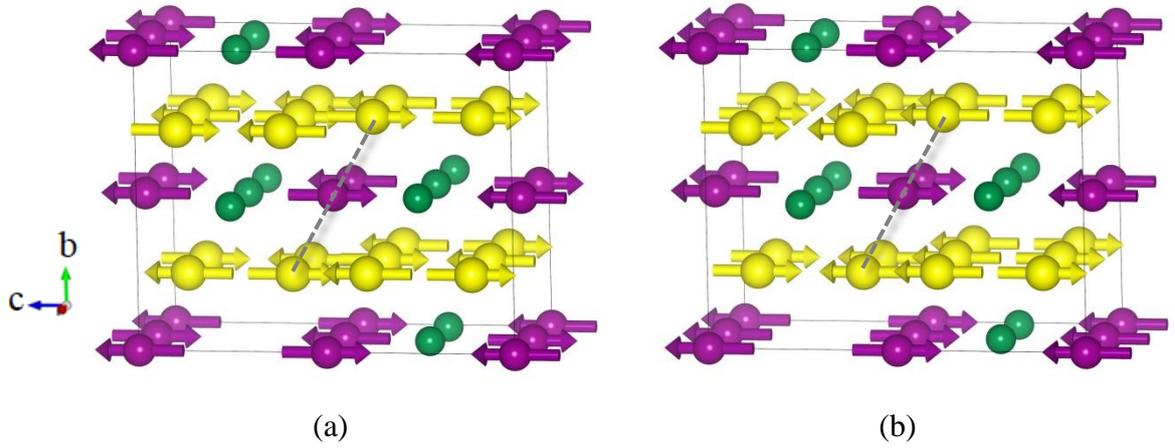

(a)  (b)

Figure S3. The orthogonal antiferromagnetic structures of CaFe$_3$O$_5$ for the AFM6 phase with propagation vector $k$ = (1/2, 0, 0) (a) and FM3 phase with $k$ = (0, 0 ,0) (b). The magnetic structure of FM3 phase is shown within doubled cell along the $a$-axis. The arrows show the directions of the Fe spins at the 4a (purple) and 8f (yellow) sites. The nonmagnetic calcium atoms at 4c sites are highlighted in green The dashed lines indicate the ferromagnetic (a) and antiferromagnetic (b) orderings within triad Fe2-Fe1-Fe2.

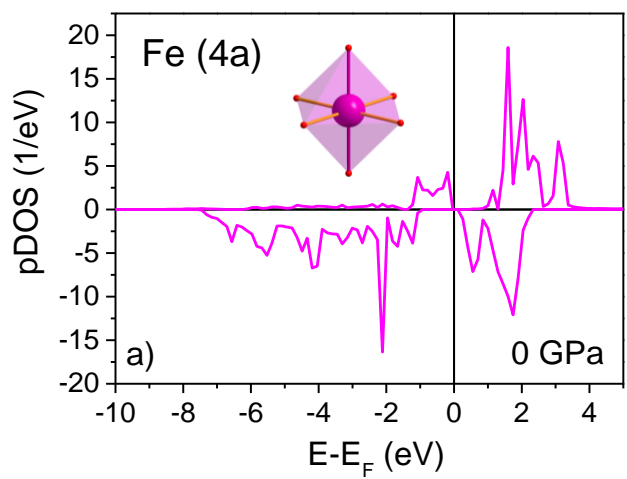
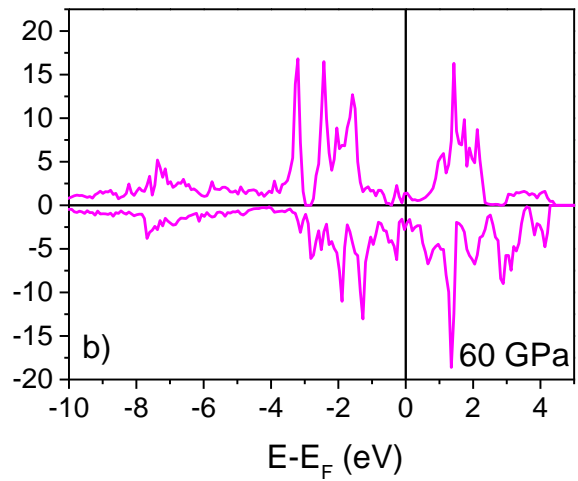
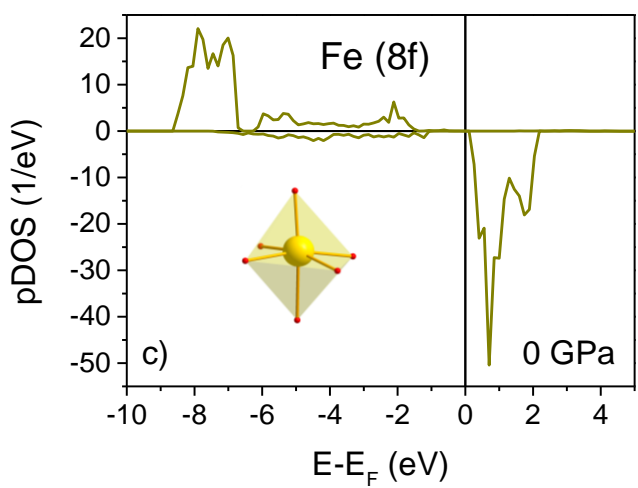
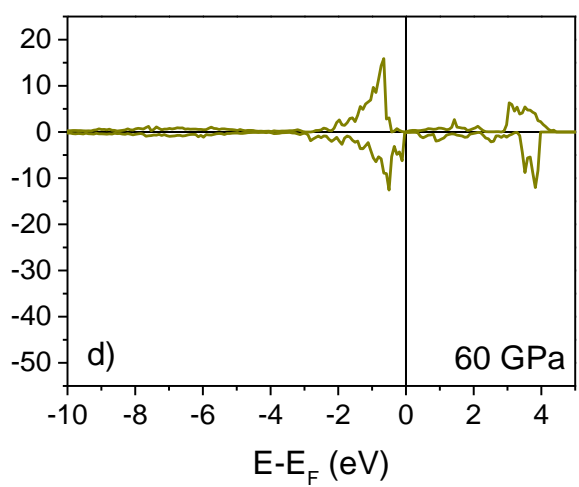
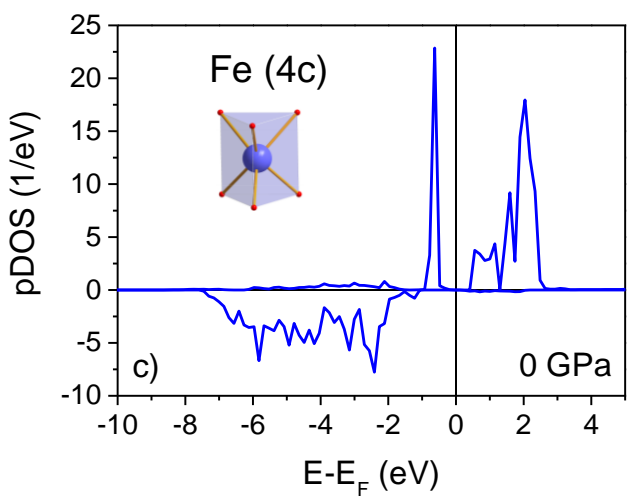
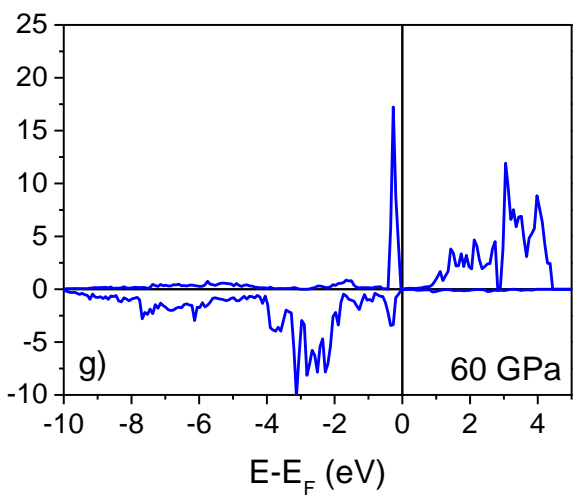

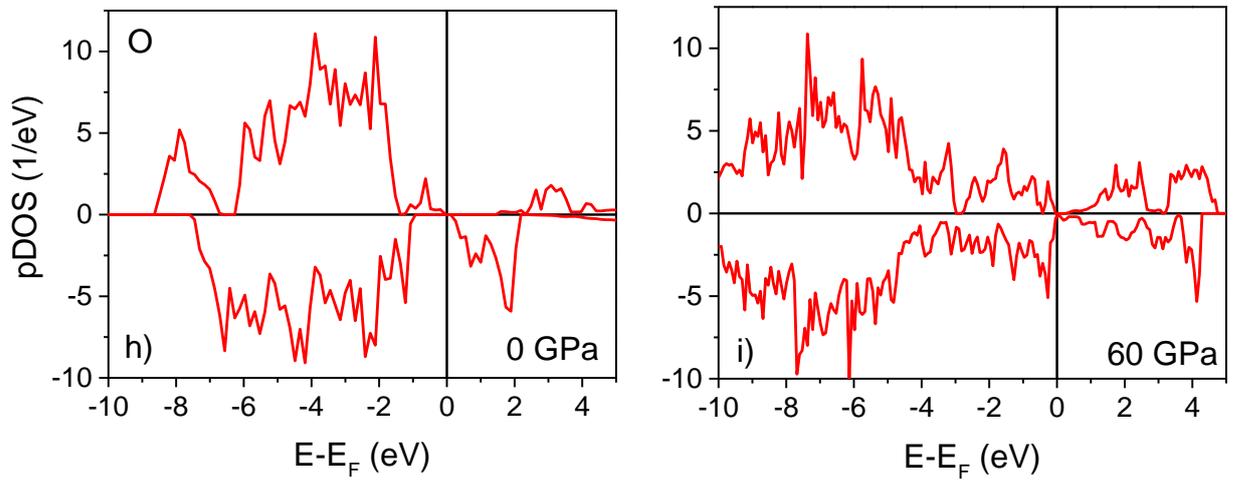

Figure S4. The iron projected density of *d*-electronic states (DOS) (a, b) Fe1, (c, d) Fe2, (e, g) Fe3; (h, i) oxygen projected density of *p*-electronic states of $Fe_4O_5$ at ambient pressure and $P_{MIT}$=60 GPa. The straight lines correspond to Fermi energy. Negative values of DOS show spin-down states.